\documentclass[english]{article}
\usepackage[T1]{fontenc}
\usepackage[latin1]{inputenc}
\usepackage{amssymb}

\makeatletter

\newcommand{\lyxaddress}[1]{
\par {\raggedright #1
\vspace{1.4em}
\noindent\par}
}

\usepackage{babel}
\makeatother
\begin{document}

\title{Unified Angular Momentum of Dyons}

\author{Shalini Dangwal, P. S. Bisht and O. P. S. Negi%
\thanks{Address from July 01 to August 31,2006:- Universität Konstanz, Fachbereich
Physik, Prof. Dr. H. Dehnen, Post fach M 677, D-78457 Konstanz, Germany.%
}}

\maketitle

\lyxaddress{\begin{center}Department of Physics,\\
Kumaun University,\\
S. S. J. Campus,\\
Almora- 263601, U.A., India\par\end{center}}

\lyxaddress{\begin{center}Email:shalini\_dangwal@rediffmail.com\\
ps\_bisht123@rediffmail.com\\
ops\_negi@yahoo.co.in\par\end{center}}

\begin{abstract}
Unified quaternionic angular momentum for the fields of dyons and
gravito-dyons has been developed and the commutation relations for
dynamical variables are obtained in compact and consistent manner.
Demonstrating the quaternion forms of unified fields of dyons (electromagnetic
fields) and gravito-dyons (gravito-Heavisidian fields of linear gravity),
corresponding quantum equations are reformulated in compact, simpler
and manifestly covariant way.
\end{abstract}
The question of existence of monopole \cite{key-1,key-2,key-3} has
become a challenging new frontier and the object of more interest
in connection with quark confinement problem of quantum chromodynamics
(QCD). The eighth decades of this century witnessed a rapid development
of the group theory and gauge field theory to establish the theoretical
existence of monopoles and to explain their group properties and symmetries.
Keeping in mind 't Hooft's solutions \cite{key-4,key-5} and the fact
that despite of the potential importance of monopoles, the formalism
necessary to describe them has been clumsy and manifestly non-covariant,
a self consistent quantum field theory of generalized electromagnetic
fields associated with dyons (particle carrying electric and magnetic
charges) has been developed \cite{key-6,key-7}. The analogy between
linear gravitational and electromagnetic fields leads the asymmetry
in Einstein's linear equation of gravity and then suggests the existence
of gravitational analogue of magnetic monopole \cite{key-8,key-9}.
Like magnetic field, Cantani \cite{key-10} introduced a new field
(called Heavisidian field) depending upon the velocities of gravitational
charges (masses) and derived the covariant equations (Maxwell's like
equations) of linear gravity. Avoiding the use of arbitrary string
variable \cite{key-1,key-2}, we have also formulated manifestly covariant
theory of gravito-dyons \cite{key-11,key-12} in terms of two four-potentials
leading to the structural symmetry between generalized electromagnetic
fields of dyons \cite{key-13,key-14} and generalised gravito-Heavisidian
fields of gravito-dyons.

In this paper, we have used quaternion analysis to combine the complex
description of dyons and gravito-dyons. Generalized fields of dyons
and gravito-dyons are combined together and the corresponding covariant
field equations and equation of motion have been derived. It has been
shown that the theory leads four different Chirality's parameters
associated respectively with electric, magnetic, gravitational and
Heavisidian charges. The study of gauge invariant and rotationally
symmetric angular momentum operator for unified quaternionic fields
of dyons and gravito-dyons has been undertaken and the commutation
relations associated with the components of angular momentum and other
dynamical parameters have been derived. The symmetry of these commutation
relations requires an additional potential term, of appropriate strength
depending upon the magnetic, gravitational and Heavisidian parameter,
besides the usual Colombian potential like term, in the Hamiltonian.
It has been shown that this unified theory reproduces the dynamics
of individual charges (masses) in the absence of others. 

In order to formulate unified theory of generalized electromagnetic
fields (associated with dyons) and generalized gravito-Heavisidian
fields (associated with gravito-dyons) of linear gravity, we describe
the properties of quaternion algebra with the use of natural units
($c=\hbar=1$), and gravitational constant is taken unity. Unified
quaternionic charge is described as \cite{key-15},

\begin{eqnarray}
Q & =(e,g,m,h) & =e-i\, g-j\, m-k\, h,\label{eq:1}\end{eqnarray}
where $e,g,m,h$ are respectively electric, magnetic, gravitational
and Heavisidian charges (masses) and $i,j,k$ are the quaternion units.
The properties of quaternion are described as under

\begin{eqnarray}
ij & =-ji= & k\,,\,\,\,\,\,\, i(ij)=(ii)j=-j\,,\nonumber \\
(ij)j=i(jj) & = & -i\,,\,\,\,\, ik=-ki=-j\,,\nonumber \\
kj & =-jk & =-i\,,\,\,\,\,\, i^{2}=j^{2}=k^{2}=-1.\label{eq:2}\end{eqnarray}
Complex structure $(e,g)$ represents the generalized charge of dyons
in electromagnetic fields while $(m,h)$ is the generalized charge
of gravito-dyons. The norm of unified quaternion charge is expressed
as,

\begin{eqnarray}
N(Q) & =Q\bar{Q} & =\bar{Q}Q=e^{2}+g^{2}+m^{2}+h^{2},\label{eq:3}\end{eqnarray}
where 

\begin{eqnarray}
\bar{Q} & = & (e,-g,-m,-h)=e+i\, g+j\, m+k\, h.\label{eq:4}\end{eqnarray}
The interaction of $a^{th}$ quaternionic charge $Q_{a}$ in the field
of $b^{th}$ quaternionic charge $Q_{b}$ depends on the quantity

\begin{eqnarray}
\bar{Q}_{a}Q_{b} & = & (W_{ab},X_{ab},Y_{ab},Z_{ab}),\label{eq:5}\end{eqnarray}
where

\begin{eqnarray}
W_{ab} & = & e_{a}e_{b}+g_{a}g_{b}+m_{a}m_{b}+h_{a}h_{b},\nonumber \\
X_{ab} & = & e_{a}g_{b}-g_{a}e_{b}+m_{a}h_{b}-h_{a}m_{b},\nonumber \\
Y_{ab} & = & e_{a}m_{b}-m_{a}e_{b}+h_{a}g_{b}-g_{a}h_{b},\nonumber \\
Z_{ab} & = & e_{a}h_{b}-h_{a}e_{b}+g_{a}m_{b}-m_{a}g_{b}.\label{eq:6}\end{eqnarray}
These four coupling parameters $W_{ab},X_{ab},Y_{ab},Z_{ab}$ may
be identified as electric, magnetic, gravitational and Heavisidian
coupling parameters associated with the basis elements $(1,i,j,k)$
of a quaternion respectively. The coupling parameter $W_{ab}$ may
also be obtained by taking the scalar product of quaternions $Q_{a}$
and $Q_{b}$ and thus equals to the norm given by equation (\ref{eq:3})
when $a=b$. It shows that if same quaternions were interacting with
each other, their behavior would be Coulomb like. $W_{ab}$ may then
be called Coulomb like coupling parameter while the second one $X_{ab}$
corresponds to the combined Chirality's parameters associated with
the interaction of two dyons (electromagnetic and gravito-dyons).
Parameter $Y_{ab}$ corresponds to the interaction between two dyons
combined by electric-gravitational and magnetic-Heavisidian charges
while the fourth one $Z_{ab}$ is the coupling parameter corresponds
to the interaction of two dyons combined by the coupling of electric-Heavisidian
and magnetic-gravitational charges. Unified quaternion valued potential
of dyons may then be defined as,

\begin{eqnarray}
V_{µ} & =(A_{µ},B_{µ},C_{µ},D_{µ}) & =A_{µ}-i\, B_{µ}-j\, C_{µ}-k\, D_{µ},\label{eq:7}\end{eqnarray}
or more generally we write as

\begin{eqnarray}
V & =(A,B,C,D) & =A-i\, B-j\, C-k\, D,\label{eq:8}\end{eqnarray}
where

\begin{eqnarray}
A & =(A_{0},A_{1},A_{2},A_{3})= & A_{0}-i\, A_{1}-j\, A_{2}-k\, A_{3},\nonumber \\
B & =(B_{0},B_{1},B_{2},B_{3})= & B_{0}-i\, B_{1}-j\, B_{2}-k\, B_{3},\nonumber \\
C & =(C_{0},C_{1},C_{2},C_{3})= & C_{0}-i\, C_{1}-j\, C_{2}-k\, C_{3},\nonumber \\
D & =(D_{0},D_{1},D_{2},D_{3})= & D_{0}-i\, D_{1}-j\, D_{2}-k\, D_{3}.\label{eq:9}\end{eqnarray}
Similarly one may define the quaternion valued unified fields as

\begin{eqnarray}
\Im_{\mu\nu} & = & F_{\mu\nu}-i\,\widetilde{F_{\mu\nu}}-j\, f_{\mu\nu}-k\,\widetilde{f_{\mu\nu}},\label{eq:10}\end{eqnarray}
where

\begin{eqnarray}
F_{\mu\nu} & = & A_{\mu,\nu}-A_{\nu,\mu}-\frac{1}{2}\varepsilon_{\mu\nu\sigma\rho}(B^{\sigma,\rho}-B^{\rho,\sigma}),\nonumber \\
\widetilde{F_{\mu\nu}} & = & B_{\mu,\nu}-B_{\nu,\mu}+\frac{1}{2}\varepsilon_{\mu\nu\sigma\rho}(A^{\sigma,\rho}-A^{\rho,\sigma}),\nonumber \\
f_{\mu\nu} & = & C_{\mu,\nu}-C_{\nu,\mu}-\frac{1}{2}\varepsilon_{\mu\nu\sigma\rho}(D^{\sigma,\rho}-D^{\rho,\sigma}),\nonumber \\
\widetilde{f_{\mu\nu}} & = & D_{\mu,\nu}-D_{\nu,\mu}+\frac{1}{2}\varepsilon_{\mu\nu\sigma\rho}(C^{\sigma,\rho}-C^{\rho,\sigma})\label{eq:11}\end{eqnarray}
and

\begin{eqnarray}
F_{\mu\nu,\nu} & = & j_{\mu}^{(e)},\,\,\,\,\,\,\,\,\,\,\,\,\,,\widetilde{F_{\mu\nu,\nu}}=j_{\mu}^{(m)},\nonumber \\
f_{\mu\nu,\nu} & = & j_{\mu}^{(G)},\,\,\,\,\,\,\,\,\,\,\,\,\,\widetilde{f_{\mu\nu,\nu}}=j_{\mu}^{(H)}.\label{eq:12}\end{eqnarray}
the (´tidle´) denotes the dual part of field tensor. In equations
(\ref{eq:11}) the field tensors $F_{\mu\nu}$, $\widetilde{F_{\mu\nu}}$
are associated with generalized fields of dyons while those for $f_{\mu\nu}$,
$\widetilde{f_{\mu\nu}}$ are associated accordingly with the generalized
gravito-Heavisidian fields of gravito-dyons \cite{key-16}. As such
the quaternion valued current may then be defined as,

\begin{eqnarray}
J_{µ} & = & j_{µ}^{(e)}-i\, j_{µ}^{(m)}-j\, j_{µ}^{(G)}-k\, j_{µ}^{(H)},\label{eq:13}\end{eqnarray}
or in general 

\begin{eqnarray}
J & = & (j^{(e)},j^{(m)},j^{(G)},j^{(H)}),\label{eq:14}\end{eqnarray}
which is the quaternion valued expression for field equation of dyons
and gravito-dyons i.e.

\begin{eqnarray}
\Im_{\mu\nu,\nu} & = & J_{\mu}.\label{eq:15}\end{eqnarray}
Lagrangian density may then be written as follows in terms of compact,
simpler and consistent quaternion notation form i.e.

\begin{eqnarray}
\mathcal{L} & = & -M-\frac{1}{8}\overline{\Im_{\mu\nu}}\,\Im_{\rho\sigma}\,\eta^{\mu\nu}\eta^{\rho\sigma}+\frac{1}{2}\overline{V_{\mu}}\, J_{\nu}\,\eta^{\mu\nu},\label{eq:16}\end{eqnarray}
where $\eta^{\mu\nu}$ is the flat space-time metric with signature
$-2$. Combined gauge covariant quaternion valued momentum may then
be written as,

\begin{eqnarray}
p\mapsto & p-\frac{1}{2}(\overline{Q}\, V+Q\,\overline{V}) & \Longrightarrow p-e\, A-g\, B-m\, C-h\, D,\label{eq:17}\end{eqnarray}
which leads to equation of motion into following compact and simpler
form \cite{key-17},

\begin{eqnarray}
M\ddot{\, x_{µ}} & = & \overline{Q}\,\Im_{\mu\nu}u^{\nu},\label{eq:18}\end{eqnarray}
where $M$ is the mass of the particle ,$\ddot{x_{µ}}$ is the acceleration
and $u^{\nu}$ is the four velocity of particle.The gauge invariant
and rotationally symmetric angular momentum in the unified quaternionic
field of dyons and gravito-dyons may be written as,

\begin{eqnarray}
\vec{J} & = & \vec{r}\times(\vec{p}-\overline{Q}\vec{\, V}\,)+(X_{ab}+Y_{ab}+Z_{ab})\,\frac{\vec{r}}{r}\:,\label{eq:19}\end{eqnarray}
where $\overline{Q}\vec{\, V}$ is the quaternionic parts. The angular
momentum given by equation (\ref{eq:19}) leads to following commutation
relations,

\begin{eqnarray}
\left[J_{k},J_{l}\right] & = & i\,\varepsilon_{klm}J_{m}\,\,(i=\sqrt{-1})\,\,\,\,(k,l,m=1,2,3),\nonumber \\
\left[J^{2},J_{k}\right] & =\left[J^{2},J_{l}\right] & =\left[J^{2},J_{m}\right]=0,\nonumber \\
\left[J_{k},r_{l}\right] & = & i\,\varepsilon_{klm}r_{m}\,,\,\,\,\,\,\,\left[\pi_{k},r_{l}\right]=-i\delta_{kl},\nonumber \\
\left[\pi_{k},\pi_{l}\right] & = & i\,(X_{ab}+Y_{ab}+Z_{ab})\varepsilon_{klm}\psi_{m},\nonumber \\
\left[J_{k},\pi_{l}\right] & =i\,\varepsilon_{klm}\pi_{m}-i & (X_{ab}+Y_{ab}+Z_{ab})r_{l}\psi_{k},\label{eq:20}\end{eqnarray}
where $\pi_{l}$ is defined as,

\begin{eqnarray}
\vec{\pi} & = & \vec{p}-\overline{Q}\,\overrightarrow{V}.\label{eq:21}\end{eqnarray}
Keeping in mind the usual Coulombian interaction problem of electric
charges, we except the following manifestly gauge invariant and rotationally
symmetric Hamiltonian for the system of unified charges i.e.

\begin{eqnarray}
\mathcal{H} & = & \frac{\pi^{2}}{2m}-\frac{W_{ab}}{r}+V(r),\label{eq:22}\end{eqnarray}
where $\vec{\pi}$ is the gauge invariant linear momentum described
by equation (\ref{eq:21}). The additional potential term $V(r)$
may be identified and decided according to the symmetry requirements
of the system. Using equation (\ref{eq:19}), the value of the operator
$J^{2}$ may be calculated as\begin{equation}
J^{2}=(\vec{r}\times\vec{\pi)}\cdot(\vec{r}\times\vec{\pi)}+(X_{ab}^{2}+Y_{ab}^{2}+Z_{ab}^{2})+2(X_{ab}+Y_{ab}+Z_{ab})\vec{r}\cdot(\vec{r}\times\vec{\pi)}\label{eq:23}\end{equation}
The third term in equation (\ref{eq:23}) is zero and hence we may
write,

\begin{eqnarray}
\frac{\pi^{2}}{2m} & = & \frac{J^{2}}{2mr^{2}}-\frac{(X_{ab}^{2}+Y_{ab}^{2}+Z_{ab}^{2})}{2mr^{2}}+\frac{p^{2}}{2m}\,\,.\label{eq:24}\end{eqnarray}
It is quite obvious that the Hamiltonian given by equation (\ref{eq:22})
possesses the higher symmetry same as that of the pure Coulomb Hamiltonian
provided the additional potential $V(r)$ in equation (\ref{eq:22})
takes the scalar form,

\begin{eqnarray}
V(r) & = & \frac{(X_{ab}^{2}+Y_{ab}^{2}+Z_{ab}^{2})}{2mr^{2}}.\label{eq:25}\end{eqnarray}
Thus the Hamiltonian given by equation (\ref{eq:22}) may be written
as,

\begin{eqnarray}
\mathcal{H} & = & \frac{\pi^{2}}{2m}-\frac{W_{ab}}{r}+\frac{(X_{ab}^{2}+Y_{ab}^{2}+Z_{ab}^{2})}{2mr^{2}},\label{eq:26}\end{eqnarray}
which leads to the following commutation relations,

\begin{eqnarray}
\left[J^{2},\mathcal{H}\right] & = & \left[J,\mathcal{H}\right]=0.\label{eq:27}\end{eqnarray}
Quaternion valued unified vector field $\overrightarrow{\psi}$ is
written as,

\begin{eqnarray}
\overrightarrow{\psi} & = & (\overrightarrow{E},\overrightarrow{M},\overrightarrow{G},\overrightarrow{H}\,),\label{eq:28}\end{eqnarray}
where $\overrightarrow{E},\overrightarrow{M},\overrightarrow{G},\overrightarrow{H}\,$
are generalized electric, magnetic, gravitational and Heavisidian
fields respectively defined \cite{key-15} as,

\begin{eqnarray}
\overrightarrow{E} & =- & \frac{\partial\overrightarrow{A}}{\partial t}-\overrightarrow{\nabla}A_{0}-\overrightarrow{\nabla}\times\overrightarrow{B},\nonumber \\
\overrightarrow{M} & =- & \frac{\partial\overrightarrow{B}}{\partial t}-\overrightarrow{\nabla}B_{0}+\overrightarrow{\nabla}\times\overrightarrow{A},\nonumber \\
\overrightarrow{G} & =- & \frac{\partial\overrightarrow{C}}{\partial t}-\overrightarrow{\nabla}C_{0}-\overrightarrow{\nabla}\times\overrightarrow{D},\nonumber \\
\overrightarrow{H} & =- & \frac{\partial\overrightarrow{D}}{\partial t}-\overrightarrow{\nabla}D_{0}+\overrightarrow{\nabla}\times\overrightarrow{C}.\label{eq:29}\end{eqnarray}
These generalized electric, magnetic, gravitational and Heavisidian
fields satisfy the following pair of Generalized Dirac Maxwell's (GDM)
equations of dyons and gravito-dyons \cite{key-11,key-12};

\begin{eqnarray}
\overrightarrow{\nabla}\cdot\overrightarrow{E} & = & j_{0}^{(e)},\,\,\,\,\,\,\,\,\overrightarrow{\nabla}\cdot\overrightarrow{M}=j_{0}^{(m)},\nonumber \\
\overrightarrow{\nabla}\cdot\overrightarrow{G} & = & j_{0}^{(G)},\,\,\,\,\,\,\,\,\,\overrightarrow{\nabla}\cdot\overrightarrow{H}=j_{0}^{(H)},\nonumber \\
\overrightarrow{\nabla}\times\overrightarrow{E} & =- & \frac{\partial\overrightarrow{M}}{\partial t}-\overrightarrow{j}^{(m)},\,\,\,\overrightarrow{\nabla}\times\overrightarrow{M}=\frac{\partial\overrightarrow{E}}{\partial t}+\overrightarrow{j}^{(e)},\nonumber \\
\overrightarrow{\nabla}\times\overrightarrow{G} & = & \frac{\partial\overrightarrow{H}}{\partial t}+\overrightarrow{j}^{(H)},\,\,\,\overrightarrow{\nabla}\times\overrightarrow{H}=-\frac{\partial\overrightarrow{G}}{\partial t}-\overrightarrow{j}^{(G)},\label{eq:30}\end{eqnarray}
where $(j_{0}^{(e)},j_{0}^{(m)},j_{0}^{(G)},j_{0}^{(H)})$ are respectively
charge densities for electric, magnetic, gravitational and Heavisidian
charges while $(\overrightarrow{j}^{(e)},\overrightarrow{j}^{(m)},\overrightarrow{j}^{(G)},\overrightarrow{j}^{(H)})$
are corresponding current source densities due to these charges. Quaternion
unified potential $V$ given by equation (\ref{eq:7}) and unified
field $\overrightarrow{\psi}$ are then related as,

\begin{eqnarray}
\overrightarrow{\psi} & = & -\frac{\partial\overrightarrow{V}}{\partial t}-\overrightarrow{\nabla}V_{0}+\overrightarrow{\nabla}\times\overrightarrow{V}.\label{eq:31}\end{eqnarray}
Operating the quaternionic differential operator $\partial=\partial_{0}-i\,\partial_{1}-j\,\partial_{2}-k\,\partial_{3}$
to equation (\ref{eq:7}) , we get the following unified form of quaternionic
potential equation,

\begin{eqnarray}
\partial\, V & = & \overrightarrow{\psi},\label{eq:32}\end{eqnarray}
which is quaternion valued unified potential equation for generalized
charges of dyons. Unified quaternion valued current may then is written
as,

\begin{eqnarray}
J & = & (j\,^{(e)},j\,^{(m)},j\,^{(G)},j\,^{(H)}),\label{eq:33}\end{eqnarray}
where

\begin{eqnarray}
\partial\,\overline{\partial}\, A & = & \square A=j\,^{(e)},\nonumber \\
\partial\,\overline{\partial}\, B & = & \square B=j\,^{(m)},\nonumber \\
\partial\,\overline{\partial}\, C & = & \square C=j\,^{(G)},\nonumber \\
\partial\,\overline{\partial}D & = & \square D=j\,^{(H)},\nonumber \\
\partial\,\overline{\partial} & = & \square=\partial_{µ}\partial_{µ}=-\frac{\partial^{2}}{\partial t^{2}}+\frac{\partial^{2}}{\partial x^{2}}+\frac{\partial^{2}}{\partial y^{2}}+\frac{\partial^{2}}{\partial z^{2}}.\label{eq:34}\end{eqnarray}
Quaternion field equation associated with unified quaternion charge
may then be written as,

\begin{eqnarray}
\partial\,\overline{\partial}\, V & =\square V & =J.\label{eq:35}\end{eqnarray}
Quaternion valued field tensor density is accordingly expressed as

\begin{eqnarray}
Q_{\mu\nu} & = & (A_{\mu\nu},B_{\mu\nu},C_{\mu\nu},D_{\mu\nu}),\label{eq:36}\end{eqnarray}
where

\begin{eqnarray}
Q_{\mu\nu} & = & V_{\mu,\nu}-V_{\nu,\mu},\nonumber \\
A_{\mu\nu} & = & A_{\mu,\nu}-A_{\nu,\mu},\nonumber \\
B_{\mu\nu} & = & B_{\mu,\nu}-B_{\nu,\mu},\nonumber \\
C_{\mu\nu} & = & C_{\mu,\nu}-C_{\nu,\mu},\nonumber \\
D_{\mu\nu} & = & D_{\mu,\nu}-D_{\nu,\mu},\label{eq:37}\end{eqnarray}
are respectively field tensors of electric, magnetic, gravitational
and Heavisidian charges and comma $(,)$ denotes partial differentiation.
$A_{µ}$ and $B_{µ}$ are dual invariant for generalized electromagnetic
fields of dyons while $C_{µ}$ and $D_{µ}$ are dual invariant under
duality transformations for generalized fields of gravito-dyons. Unified
quaternion valued current $J_{µ}$ and field tensor $Q_{\mu\nu}$
are then related in the following manner;

\begin{eqnarray}
Q_{\mu\nu,\nu} & = & J_{µ},\label{eq:38}\end{eqnarray}
which is equivalent to the following quaternionic form;

\begin{eqnarray}
\overline{\partial}\,\psi & = & J.\label{eq:39}\end{eqnarray}
as such, an unified field theory of generalised electromagnetic and
Heavisidian fields have been developed in view of the fact that these
fields essentially possess the structural symmetry. Quaternion charge
defined by equation (\ref{eq:1}) represents the theory of particles
carrying simultaneously electric, magnetic, gravitational and Heavisidian
charges. Equation (\ref{eq:1}) has been described as the combination
of two complex charges of dyons $(e,g)$ and gravito-dyons $(m,h)$.
Equation (\ref{eq:6}) represents the coupling parameters $W_{ab}$
, $X_{ab}$ , $Y_{ab}$ and $Z_{ab}$ . This equation shows that in
the presence of only electric charge the coupling becomes $W_{ab}=e_{a}e_{b}$
while the other parameters $X_{ab},Y_{ab}$ and $Z_{ab}$vanish. If
the particles are considered as dyons $(e,g)$ of electromagnetic
fields, we have $W_{ab}=e_{a}e_{b}+g_{a}g_{b}$ and $X_{ab}=e_{a}g_{b}-g_{a}e_{b}$
which are respectively named as electric and magnetic coupling parameters.
For gravito-dyons $(m,h)$ we have only $W_{ab}=m_{a}m_{b}+h_{a}h_{b}$
and $X_{ab}=m_{a}h_{b}-h_{a}m_{b}$ . For other dyon like $(e,h)$
i.e. electric and Heavisidian charges we get $W_{ab}=e_{a}e_{b}+h_{a}h_{b}$
and $Z_{ab}=e_{a}h_{b}-h_{a}e_{b}$ . For other dyon like $(g,m)$
i.e. gravitational charge and magnetic monopole, we get $W_{ab}=m_{a}m_{b}+g_{a}g_{b}$
and $Z_{ab}=g_{a}m_{b}-m_{a}g_{b}$. For other dyon like $(e,m)$
i.e. electric charge and gravitational charge (mass), we get $W_{ab}=e_{a}e_{b}+m_{a}m_{b}$
and $Y_{ab}=e_{a}m_{b}-m_{a}e_{b}$. Similarly for purely hypothetical
dyons $(h,g)$ i.e. Heavisidian and magnetic charges we get $W_{ab}=h_{a}h_{b}+g_{a}g_{b}$
and $Y_{ab}=h_{a}m_{b}-m_{a}h_{b}$. As such, the unified theory of
quaternionic charge is the combined theory of fields characterizing
the study of fields associated with various types of dyons.

Equation (\ref{eq:7}) represents unified quaternion valued potential
of dyons. It is to be noted that here we have combined four-charges
(electric, magnetic, gravitational and Heavisidian) in terms of tetrad
combination in the context of special relativity (not in terms of
generalised coordinates) and maintained the abelian gauge structure
in terms of four photons with a the group is defined as $\mathcal{G=\mathrm{U^{(e)}(1)\times}\mathrm{U^{(m)}(1)\times\mathrm{U^{(G)}(1)\times\mathrm{U^{(H)}(1)}}}}$.
While the quaternion structure leads to rotation in charge space and
it is our next target to find out these investigations of quaternion
rotation shortly to mix up the gauge structures. here we have the
possibility to consider the gravitational charge $m$ both positive
or negative. This should imply that gravitational and Heavisidian
charges may be either positive or negative dealing with the concept
of negative masses. On the other hand, properties of quanternion algebra
are incorporated to derive the field equation (\ref{eq:10}) as a
quaternion and the coefficients of its basis elements represent the
algebraic gauge structure field equations. Lagrangian density (\ref{eq:16})
has been shown to yield the compact and simpler representation of
combined dyonic field equation (\ref{eq:15}) and equation of motion
(\ref{eq:18}). The gauge invariant and rotationally symmetric angular
momentum in the unified quaternionic fields of dyons and gravito-dyons
given by equation (\ref{eq:19}) contains an extra term $J_{res}=(X_{ab}+Y_{ab}+Z_{ab})\frac{\overrightarrow{r}}{r}$
of residual angular momentum in addition to the usual kinetic angular
momentum $J_{kin}=\vec{r}\times(\vec{p}-\overline{Q}\vec{\, V}\,)$.
The rotationally symmetric nature of angular momentum is also reflected
in the commutation relation given by equation (\ref{eq:20}). The
commutation relations given by equations (\ref{eq:20}) and (\ref{eq:27})
show that the operator and are the constant of motion. The commutation
relation (\ref{eq:27}) demands an additional term in the Hamiltonian
so that it consists the higher symmetry as the pure Coulomb Hamiltonian.
From the above analysis it may be concluded that besides the potential
importance of monopole as intrinsic part of grand unified theories,
monopoles and dyons may provide even more ambitious model to purport
the unification of gravitation with strong and electro-weak forces.The
electric, magnetic, gravitational and Heavisidian fields defined by
equation (\ref{eq:29}) describe the pair of Maxwell's equations associated
with generalized fields of dyons $(e,g)$ and gravito-dyons $(m,h)$.
Equation (\ref{eq:39}) describes the unified forms of quaternion
four-current density and it may be shown that these quaternion field
equations are invariant under quaternion and duality transformations.

\textbf{Acknowledgment}: - O. P. S. Negi is thankful to Prof. Dr.
H. Dehnen, Universität Konstanz, Fachbereich Physik , Post fach M
677, D-78457 Konstanz, Germany for his hospitality at Konstanz. He
is also grateful to German Academic Exchange Service (DAAD) for financial
support under re-invitation fellowship programme.

\end{document}